\documentclass{sf2a-conf}
\usepackage{graphicx}
\usepackage{natbib}
\begin{document}

\TitreGlobal{SF2A 2008}

\title{The magnetic field of solar prominences.}
\author{Paletou, F.}\address{Laboratoire d'Astrophysique de Toulouse-Tarbes, Universit\'e de Toulouse, CNRS, 14 ave. E. Belin, 31400 Toulouse}
\runningtitle{The magnetic field of solar prominences}
\setcounter{page}{237}

\index{Paletou, F.}

\maketitle
\begin{abstract}
  In his famous monographs, Einar Tandberg-Hanssen writes that ``the
  single, physically most important parameter to study in prominences
  may be the magnetic field. Shapes, motions, and in fact the very
  existence of prominences depend on the nature of the magnetic field
  threading the prominence plasma''. Hereafter we sumarize recent
  contributions and advances in our knowledge about the magnetic field
  of solar prominences. It mostly relies on high resolution and high
  sensitivity spectropolarimetry made both in the visible and in the
  near infrared.
\end{abstract}
%
\section{Introduction}

Solar prominences (filaments) are made-up of dense and cool
chromospheric plasma hanging in the hot and low density corona
(Tandberg-Hanssen 1995). Besides its intrinsic interest, as a natural
laboratory for plasma physics, the study of these structures is also
of a more general interest in the frame of space weather
studies. Indeed, among other closed magnetic regions such as active
regions, eruptive prominences are often associated with coronal mass
ejections, or CMEs, that is huge plasma ``bubbles'' ejected from the
solar corona and able to strongly affect Sun-Earth relationships, by
their interactions with the terrestrial magnetosphere (see e.g.,
Gopalswamy et al. 2006, for a recent review upon the various
precursors of CMEs).

Despite systematic observations made since the nineteenth century and
decades of study, prominence formation mechanisms are still not well
understood. In particular, yet no theory can fully explain their
remarkable stability in a hotter and less dense medium. However, since
the plasma $\beta$ is low in prominences, the magnetic field is very
likely to play a major role in the physical scenarios which could
explain prominences formation, stability and, finally, the triggering
of these instabilities leading to CMEs (see Fig. 1).

However, the 3D magnetic field topology of solar prominences is not
\emph{directly} measureable in the corona. Even though indirect
methods are available, the best possible determination of prominences
magnetic fields comes from the inversion of spectropolarimetric data,
which collection still remains a difficult task. He\,{\sc i}
multiplets such as $\lambda 10830$~\AA\ in the near-infrared, and the
Fraunhofer ``yellow line'' $D_{3}$ at $\lambda 5876$~\AA~are the best
tools, so far, to study prominence magnetic fields. Indeed, only a few
spectral lines are intense enough for ground-based observations in the
optical spectrum of solar prominences i.e., at these wavelength at
which spectropolarimetry is usually done. These helium multiplets
provide, even if they are fainter than H$\alpha$ for instance, the
most suitable information necessary for the purpose of determining the
magnetic field pervading the prominence plasma. Indeed, H$\alpha$ is
generally optically thick in prominences which, together with its
hyperfine atomic structure, makes this spectral line still much more
difficult to deal with, as compared to the above-mentioned helium
multiplets. First spectropolarimetric observations of prominences and
associated results about the magnetic field properties have been
recently reviewed by Paletou \& Aulanier (2003).

\begin{figure}[t]
   \centering
   \includegraphics[width=10cm]{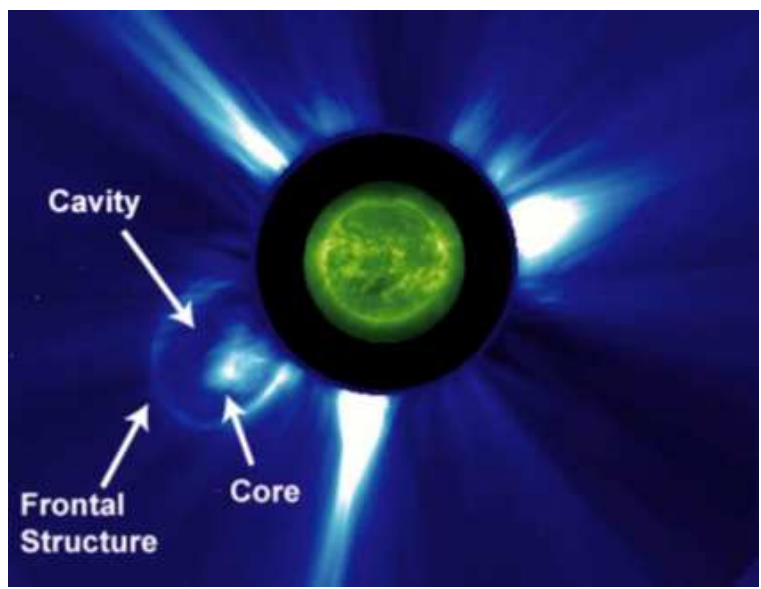}
   \caption{A view of the standard three-part structure of a coronal
     mass ejection, as observed on December 20, 2001 with the LASCO
     coronagraph on-board SoHO. The so-called bright core is the
     remains of an eruptive prominence (from Gopalswamy et al. 2006).}
   \label{fig1}
 \end{figure}

\section{Recent advances}

After very fruitful years of observations made mostly in the 80's,
mainly at the \emph{Pic du Midi} in France and at Sacramento Peak by
NSO and HAO groups in the USA, spectropolarimetry of prominences seems
to resume after the pionnering work of Lin et al. (1998) and the
full-Stokes observations of a filament (i.e., a prominence as seen on
the disk) in the $\lambda 10830$~\AA\ multiplet.

A few years after, the first \emph{full-Stokes} and high spectral
resolution observation of a prominence in the He\,{\sc i} $D_{3}$
multiplet is made at TH\'eMIS (Paletou et al. 2001). These
observations revealed a mixture of Hanle and Zeeman effects
signatures. And the direct analysis, under the weak-field
approximation, of the measured Stokes $V$ signals pointed at a
longitudinal magnetic field value of the order of 40 G, i.e. quite
larger than what was usually measured in quiescent prominences (Leroy
1989, Leroy et al. 1984). Magnetic field strengths of the order of 50 G
were also reported by Wiehr \& Bianda (2003).

The new TH\'eMIS observations have also led to a revision of magnetic
field inversion tools (L\'opez Ariste \& Casini 2002). In particular,
these authors demonstrated how taking into account \emph{all} Stokes
parameters, and not only linear polarization signals, can increase the
reliability of the inversion process.

Shortly after, Casini et al. (2003) published the first \emph{maps} of
the vector magnetic field i.e., its modulus, azimuth and inclination,
as inferred from \emph{Advanced Stokes Polarimeter} observations made
at D$_3$ of He\,{\sc i}, at the Dunn Solar Tower (DST, NSO/SP, USA). Even
stronger field measurements were confirmed, up to 70 G, and variations
of the orientation of the magnetic field across the prominence body
were revealed.

\subsection{Indirect methods}

Long sequences of observations, more likely provided by spaceborne
observatories, offer the possibility, to a certain extent, for an
indirect diagnosis of the mean magnetic field in a
filament/prominence. Using a 7h30 observing sequence made with the CDS
EUV spectrometer on-board SoHO, and from the identification of certain
oscillation modes associated to Alfv\'en and magnetoacoustic waves,
R\'egnier et al. (2001) could infer the angle between the mean
magnetic field and the filament long axis, and the magnetic field
strength vs. the electronic density (although the latter remained
undetermined).

Some important properties of the magnetic field of prominences can
also be deduced from vector magnetic maps made at the
\emph{photospheric} level, in association with H$\alpha$ imagery of
the area below were stands the prominence. The latter images have
to be used \emph{simultaneously} with the photospheric vector
magnetograms, and analysed taking into account the so-called
\emph{chirality rules} which were established by Martin (1998).

Taking advantage of the multi-line capabilities of TH\'eMIS, and after
a delicate analysis of several sets of data, L\'opez Ariste et
al. (2006) could indeed identify the presence of photospheric magnetic
field dips, also known as \emph{bald patches} (see Fig. 2). According
to these authors, this observed magnetic field topology in the
photosphere tends to support MHD models of prominences based on
magnetic dips located within weakly twisted flux tubes.

\subsection{Near-infrared observations}

The $\lambda 10830$~\AA\,multiplet of He\,{\sc i} is routinely
observed at the German VTT with the TIP polarimeter developed at the
IAC (M\'artinez Pillet et al. 1999, Collados et al. 2007). Filaments
and prominences observations made with such instruments have recently
conducted to very interesting results.

From observations of a filament at disk center, Trujillo Bueno et
al. (2002) could show, from the analysis of linear polarization
signals observed in the two well-separated components of the $\lambda
10830$~\AA\,multiplet, and taking advantage of the forward-scattering
geometry of this observation, that the effect of selective absorption
from the ground-level of the triplet system of He\,{\sc i} is at
work. This implies the presence of magnetic fields of the order of a
few gauss that are highly inclined with respect to the solar radius
vector.

\begin{figure}[t]
   \centering
   \includegraphics[width=10cm]{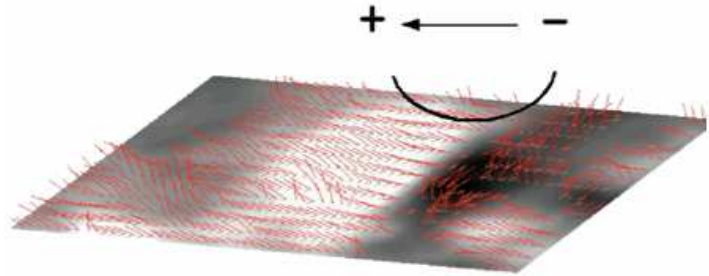}
   \caption{The presence of magnetic dips supporting the prominence
     plasma against gravity can be inferred from a careful analysis of
     both (photospheric) vector magnetic field maps and simultaneous
     H$\alpha$ images of the filament and its environment (from
     L\'opez Ariste et al. 2006, using TH\'eMIS multi-wavelength
     observations).}
   \label{fig2}
 \end{figure}

More recently, Merenda et al. (2006) published a quite surprising
result concerning the orientation of the magnetic field deduced from
the observation of a \emph{polar crown} prominence. A magnetic field of 30 G
strength inclined by about 25$^{\rm o}$ with respect to the local
solar vertical direction was inferred. These authors could also deduce
from their analysis that, this nearly vertical magnetic field appeared
to be slightly rotating around a fixed direction in space as one
proceeds along the direction of the spectrograph's slit (which was, in
that case, parallel to the local limb).

\section{TH\'eMIS on the front line}

Nowadays, the 1-m aperture class TH\'eMIS solar telescope installed at
the \emph{Observatorio del Teide} in Iza\~na (Tenerife, Spain) is the
tool of choice for observing programmes dedicated to the
spectropolarimetry of solar prominences. Since 2006, and the
rejuvenation of the pool of detectors for the MTR observing mode (see
e.g., Paletou \& Molodij 2001), it provides indeed a \emph{unique}
capability of high spectral resolution, multi-line spectropolarimetric
observations \emph{simultaneously} in the visible and in the
near-infrared spectral domains.

On Fig.~3, we display Stokes profiles extracted from data taken on
June 2007. With our MTR setup, observations of D$_3$, H$\alpha$ and
$\lambda 10830$~\AA\, spectral domains are made simultaneously. For
these observations, a polarimetric sensitivity better than $10^{-4}$
was reached. Such a combination of measurements with high spectral
resolution and polarimetric sensitivity is, so far, a unique
capability which is fully relevant for the deeper study of prominences
magnetic fields in the coming years.

In the frame of our programme of spectropolarimetric observation of
prominences, we could also put in evidence ``enigmatic'' circular
polarization signals in H$\alpha$, both symmetric and having
amplitudes which can be comparable to linear polarization signals,
unlike what is predicted by the theory of the Hanle effect (see e.g.,
Landi Degl'Innocenti 1982). This was confirmed by observations
made at the DST with the ASP spectropolarimeter (L\'opez Ariste et
al. 2005). Even though other groups, such as Stenflo's (ETH, Z\"urich)
using the 45-cm aperture Gregory-Coud\'e telescope at Locarno with the
ZImPol spectropolarimeter could not yet confirm our findings (Ramelli
et al. 2006), the analysis of data collected during our 2007 and 2008
observing campaigns do confirm the existence of such $V$ signals. Our
present study of this set of data aims at understanding under which
conditions such circular polarization signals appear and, which are
the physical implications of their presence or absence.

There are several physical processes capable of generating the
observed net circular polarization (hereafter NCP) at H$\alpha$ and,
it is still unclear which one, or which combination of effects, is
indeed at work in prominences. This may result from the presence of
\emph{electric} fields in the prominence plasma (see e.g., Casini \&
Manso Sainz 2006). However, some authors have also shown that
collisional effects, either anisotropic (Derouich 2007) or isotropic
(\v St\v ep\'an \& Sahal-Br\'echot, these proceedings), can also
generate NCP.

\begin{figure}[t]
  \begin{minipage}[b]{0.40\linewidth}
      \centering \includegraphics[width=9cm]{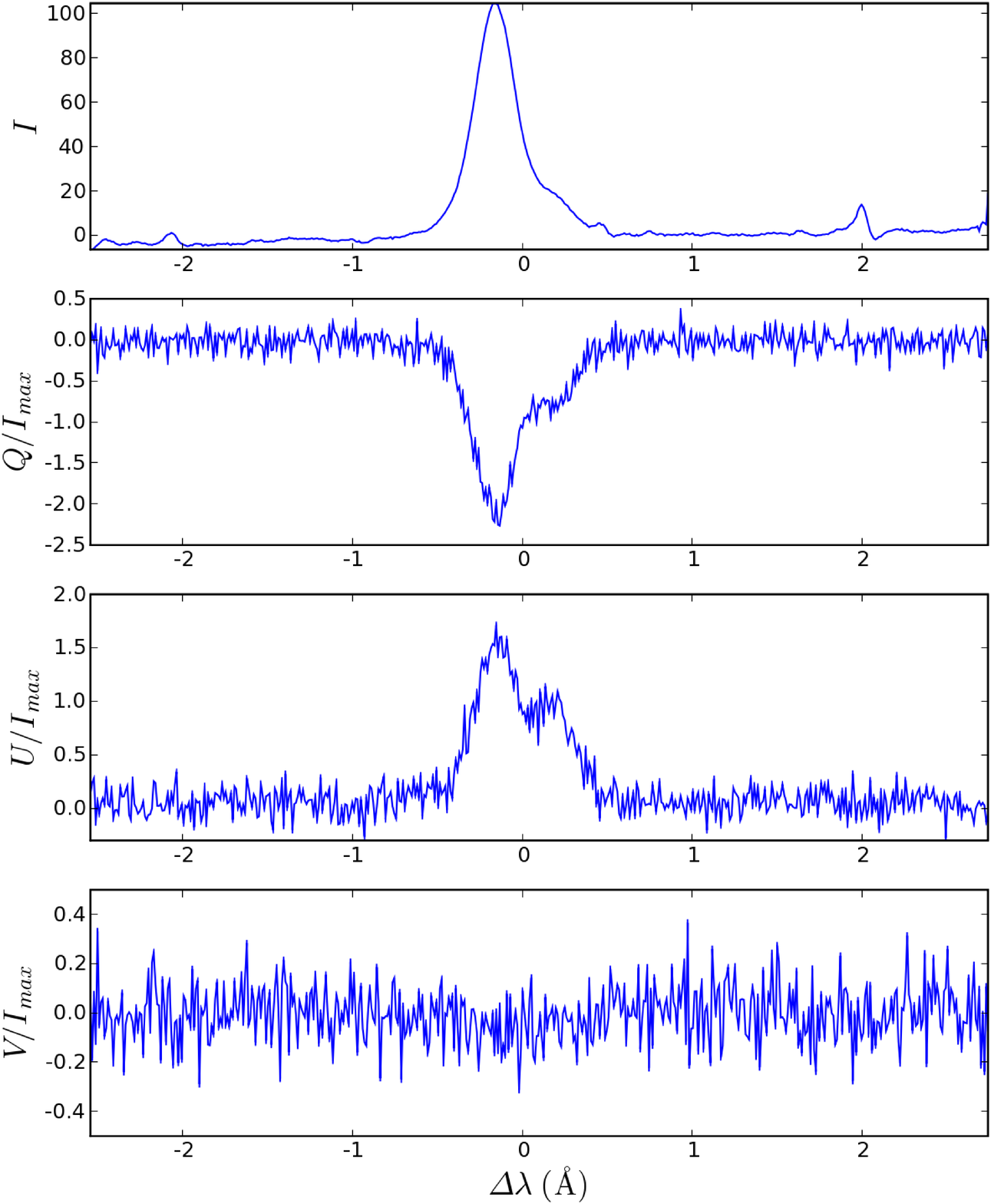}
   \end{minipage}\hfill
   \begin{minipage}[b]{0.48\linewidth}   
      \centering \includegraphics[width=9cm]{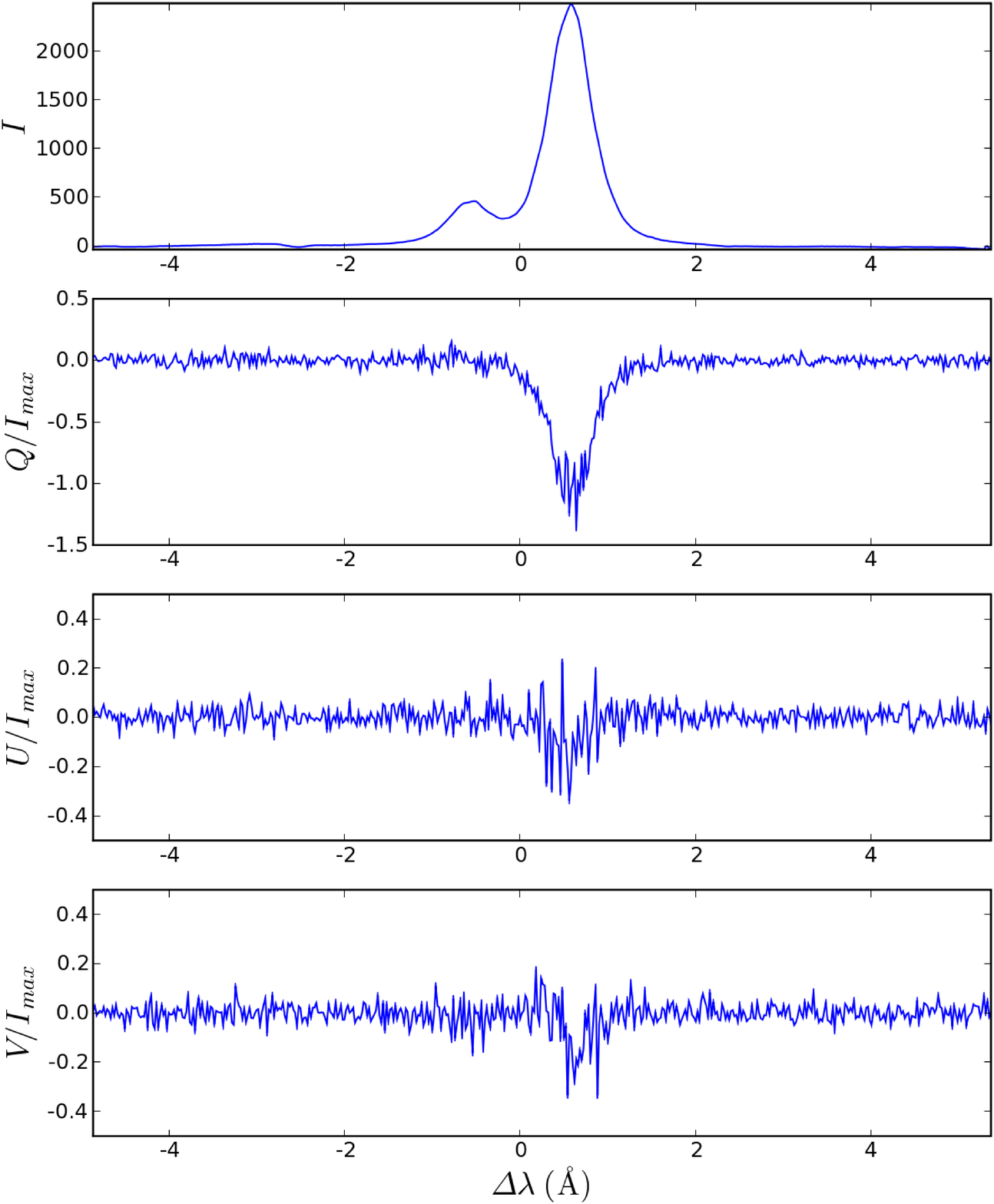}
   \end{minipage}
   \caption{Full-Stokes measurements of polarized signals formed in a
     solar prominence observed simultaneously with TH\'eMIS on June
     2007, at 5876 \AA\,(left) and at 10830 \AA\,(right). The Stokes
     profiles $Q$, $U$ and $V$ are normalized to the maximum of $I$
     after removal of the scattered light. H$\alpha$ was also observed
     simultaneously with our MTR set-up. Polarization signals
     displayed here have been obtained with a sensitivity better than
     $10^{-3}$.}
\end{figure}

\section{The need for complex radiative modelling}

It happens that measurements of the ratio between the amplitude of the
two components of Stokes $I$, resulting from the atomic fine
structure, for the He\,{\sc i} $\lambda 10830$~\AA\, and $D_{3}$
multiplets (see e.g., Fig. 11 in L\'opez Ariste \& Casini 2002) are
often in contradiction with the commonly used hypothesis of
\emph{optically thin} multiplets (see e.g., Bommier 1977).
 
Besides, up to now the most recent radiative models (Labrosse \&
Gouttebroze 2001, 2004) still assume mono-dimensional (1D) static
slabs and \emph{no} atomic fine structure for the He\,{\sc i}
model-atom, which lead to the synthesis of unrealistic Gaussian
profiles. Given the high spectral resolution of actual observations,
it is therefore important to use the best numerical radiative
modelling tools in 2D geometry, as a first step, and a more detailed
He\,{\sc i} atomic model in order to improve our spectral diagnosis
capability.

As a first application of the new 2D radiative transfer code developed
by us (Paletou \& L\'eger 2007, L\'eger et al. 2007), we have shown
how \emph{multi-thread} models, for which one considers the emission
resulting from a bunch of cool small-scale structures distributed
along the line of sight, could explain the measured intensity ratios
(L\'eger \& Paletou 2008).

Such a forward complex radiative modelling now have to be exploited
and developed further. It should also be used for the generation of
synthetic polarization signals, possibly combined with recently
developed numerical tools such as {\sc hazel} (Asensio Ramos et
al. 2008).

\section{Conclusions}

The spectropolarimetry of solar prominences have been renewed during
the last decade with the advent of both new telescopes, new
spectropolarimeters and detectors. Even though the data collection has
not been huge so far, most of the measurements of quality have led to
new and surprising results.

In that field, the solar telescope TH\'eMIS with its unique multi-line
spectropolarimetric capability is definitely, at the present time, the
instrument of choice for such studies. In a near-future, only the
dedicated {\sc ProMag} spectropolarimeter currently developed at
HAO/NCAR, and to be deployed at the Evans Solar Facility (NSO/SP,
USA), will provide almost comparable sets of data.

It is finally expected that the future 4-m aperture solar telescope
EST will allow too, at the 2020 horizon, for the collection of the
most suitable combination of spectropolarimetric data necessary for
ever more precise determinations of the magnetic fields of solar
prominences.

\begin{acknowledgements}
  Arturo L\'opez Ariste (TH\'eMIS, CNRS), Roberto Casini (HAO, NCAR,
  Boulder), Reza Rezai (KIS, Freiburg) and Ludovick L\'eger (LATT,
  U. Toulouse, CNRS) are the main collaborators of our observing
  programme related to the spectropolarimetry of solar
  prominences. TH\'eMIS is operated on the Island of Tenerife by
  CNRS-CNR in the Spanish \emph{Observatorio del Teide} of the
  \emph{Instituto de Astrof\'{\i}sica de Canarias}.
\end{acknowledgements}

\end{document}